%% file: symmetric.tex
\input harvmac

\input youngtab

 \def\quad{{\ \ }}

\def\la{\langle}
\def\ra{\rangle}

\def\vac{{\rm vac}}

\let\includefigures=\iftrue
\newfam\black
\includefigures
\input epsf
\def\figin{\epsfcheck\figin}\def\figins{\epsfcheck\figins}
\def\epsfcheck{\ifx\epsfbox\UnDeFiNeD
\message{(NO epsf.tex, FIGURES WILL BE IGNORED)}
\gdef\figin##1{\vskip2in}\gdef\figins##1{\hskip.5in}
\else\message{(FIGURES WILL BE INCLUDED)}%
\gdef\figin##1{##1}\gdef\figins##1{##1}\fi}
\def\DefWarn#1{}
\def\figinsert{\goodbreak\midinsert}
\def\ifig#1#2#3{\DefWarn#1\xdef#1{fig.~\the\figno}
\writedef{#1\leftbracket fig.\noexpand~\the\figno}%
\figinsert\figin{\centerline{#3}}\medskip\centerline{\vbox{\baselineskip12pt
\advance\hsize by -1truein\noindent\footnotefont{\bf Fig.~\the\figno:}
#2}}
\bigskip\endinsert\global\advance\figno by1}
\else
\def\ifig#1#2#3{\xdef#1{fig.~\the\figno}
\writedef{#1\leftbracket fig.\noexpand~\the\figno}%
#2}}
\global\advance\figno by1}
\fi


\def\sym{  \> {\vcenter  {\vbox
                  {\hrule height.6pt
                   \hbox {\vrule width.6pt  height5pt
                          \kern5pt
                          \vrule width.6pt  height5pt
                          \kern5pt
                          \vrule width.6pt height5pt}
                   \hrule height.6pt}
                             }
                  } \>
               }
\def\fund{  \> {\vcenter  {\vbox
                  {\hrule height.6pt
                   \hbox {\vrule width.6pt  height5pt
                          \kern5pt
                          \vrule width.6pt  height5pt }
                   \hrule height.6pt}
                             }
                       } \>
               }
\def\anti{ \>  {\vcenter  {\vbox
                  {\hrule height.6pt
                   \hbox {\vrule width.6pt  height5pt
                          \kern5pt
                          \vrule width.6pt  height5pt }
                   \hrule height.6pt
                   \hbox {\vrule width.6pt  height5pt
                          \kern5pt
                          \vrule width.6pt  height5pt }
                   \hrule height.6pt}
                             }
                  } \>
               }

\Title{\vbox{\baselineskip12pt\hbox{hep-th/0612022}}}
{\vbox{\centerline{Wilson Loops as $D3$-Branes}}}

\centerline{Jaume Gomis\foot{jgomis@perimeterinstitute.ca} and Filippo Passerini\foot{fpasserini@perimeterinstitute.ca}}
\medskip\medskip

\bigskip\centerline{\it Perimeter Institute for Theoretical Physics}
\centerline{\it Waterloo, Ontario N2L 2Y5, Canada$^{1,2}$}
\vskip .05in

\bigskip\centerline{\it Department of Physics and Astronomy}
\centerline{\it University of Waterloo,  Ontario N2L 3G1, Canada$^2$}
\vskip .2in
\centerline{Abstract}

We prove that  the half-BPS Wilson loop operator  of ${\cal N}=4$ SYM
in a symmetric representation of the gauge group has a bulk
gravitational description in terms of a single $D3$-brane in AdS$_5\times$S$^5$, as argued in hep-th/0604007.
We also show that a half-BPS Wilson loop operator in an arbitrary representation is described by the $D3$-brane
configuration proposed in hep-th/0604007. This is demonstrated  by explicitly   integrating out the degrees of freedom on the $D3$-branes and showing that they insert a half-BPS Wilson loop operator into the ${\cal N}=4$ SYM path integral in the desired representation.

\Date{12/2006}

\newsec{Introduction}

Recently, the bulk description of all half-BPS Wilson loop operators of ${\cal N}=4$ SYM has been given 
\lref\GomisSB{
  J.~Gomis and F.~Passerini,
   ``Holographic Wilson loops,''
  JHEP {\bf 0608}, 074 (2006)
  [arXiv:hep-th/0604007].
}
\GomisSB\ in terms of D-branes in AdS$_5\times$S$^5$ 
(see   
\lref\DrukkerKX{
  N.~Drukker and B.~Fiol,
  ``All-genus calculation of Wilson loops using D-branes,''
  JHEP {\bf 0502}, 010 (2005)
  [arXiv:hep-th/0501109].
}
\lref\YamaguchiTE{
  S.~Yamaguchi,\hskip-1pt
  ``Bubbling geometries for half BPS Wilson lines,''\hskip-1pt
  arXiv:hep-th/0601089.
}
\DrukkerKX\YamaguchiTE\ for previous work). A half-BPS Wilson loop -- labeled by a representation of $U(N)$ with Young tableau --
\medskip
 \ifig\Youngtabcolumnsrows{A Young tableau. For $U(N)$ $P\leq N$
and $M$ is  arbitrary.}{\epsfxsize2in\epsfbox{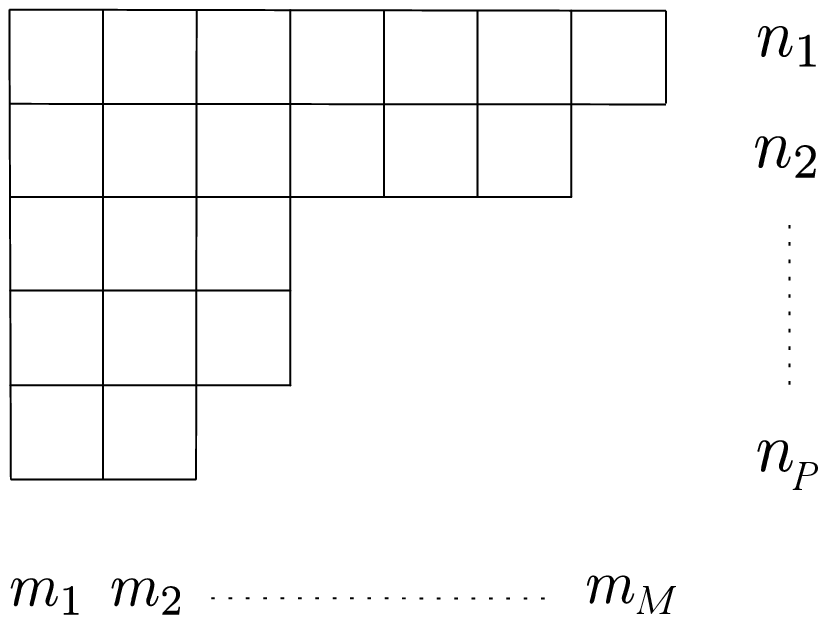}}
\noindent
 can be described \GomisSB\   in terms of $M$ $D5$-branes or alternatively in terms of $P$ $D3$-branes in AdS$_5\times$S$^5$. This 
 generalizes the bulk description of a Wilson loop in the fundamental representation
 \lref\ReyIK{
  S.~J.~Rey and J.~T.~Yee,
  ``Macroscopic strings as heavy quarks in large N gauge theory and  anti-de
  Sitter supergravity,''
  Eur.\ Phys.\ J.\ C {\bf 22}, 379 (2001)
  [arXiv:hep-th/9803001].
}
 \lref\MaldacenaIM{
  J.~M.~Maldacena,
  ``Wilson loops in large N field theories,''
  Phys.\ Rev.\ Lett.\  {\bf 80}, 4859 (1998)
  [arXiv:hep-th/9803002].
}
 in terms of a   string worldsheet \ReyIK\MaldacenaIM\ to all other representations.  For other work see e.g.
 \nref\YamaguchiTQ{
  S.~Yamaguchi,
   ``Wilson loops of anti-symmetric representation and D5-branes,''
  JHEP {\bf 0605}, 037 (2006)
  [arXiv:hep-th/0603208].
}
\nref\HartnollHR{
  S.~A.~Hartnoll and S.~Prem Kumar,
  ``Multiply wound Polyakov loops at strong coupling,''
  Phys.\ Rev.\ D {\bf 74}, 026001 (2006)
  [arXiv:hep-th/0603190].
}
\nref\LuninXR{
  O.~Lunin,
   ``On gravitational description of Wilson lines,''
  JHEP {\bf 0606}, 026 (2006)
  [arXiv:hep-th/0604133].
} 
 \lref\GomisCU{
  J.~Gomis and C.~Romelsberger,
   ``Bubbling defect CFT's,''
  JHEP {\bf 0608}, 050 (2006)
  [arXiv:hep-th/0604155].
}
\nref\OkuyamaJC{
  K.~Okuyama and G.~W.~Semenoff,
   ``Wilson loops in N = 4 SYM and fermion droplets,''
  JHEP {\bf 0606}, 057 (2006)
  [arXiv:hep-th/0604209].
} 
\nref\HartnollIS{
  S.~A.~Hartnoll and S.~P.~Kumar,
   ``Higher rank Wilson loops from a matrix model,''
  JHEP {\bf 0608}, 026 (2006)
  [arXiv:hep-th/0605027].
} 
 \nref\GiombiDE{
  S.~Giombi, R.~Ricci and D.~Trancanelli,
  ``Operator product expansion of higher rank Wilson loops from D-branes and
  matrix models,''
  JHEP {\bf 0610}, 045 (2006)
  [arXiv:hep-th/0608077].
}
 \nref\TaiBT{
  T.~S.~Tai and S.~Yamaguchi,
   ``Correlator of fundamental and anti-symmetric Wilson loops in AdS/CFT
  correspondence,''
  arXiv:hep-th/0610275.
}
\lref\GomisKB{
  J.~Gomis, S.~Moriyama and J.~w.~Park,
  ``Open + closed string field theory from gauge fields,''
  Nucl.\ Phys.\ B {\bf 678}, 101 (2004)
  [arXiv:hep-th/0305264].
}
\noindent \hskip-30pt
\YamaguchiTQ-\TaiBT.
 
 In \GomisSB, it was argued that the   $D3_k$-brane\foot{$k$ is the   fundamental string charge  dissolved on the brane.} solution  in AdS$_5\times$S$^5$ of \ReyIK\DrukkerKX\ corresponds to a half-BPS Wilson loop operator in the $k$-th symmetric representation, while a Wilson loop operator in the representation given by Fig. 1  corresponds to the array of $P$ branes $(D3_{n_1},D3_{n_2},\ldots,D3_{n_P})$.

In this note we give a first principles derivation of this proposal. We complete the analysis in \GomisSB\ by studying the flat space  brane configuration which yields in the near horizon limit the  $P$ $D3$-branes $(D3_{n_1},D3_{n_2},\ldots,D3_{n_P})$ in AdS$_5\times$S$^5$ dual to the Wilson loop operator labeled by Fig. 1. This  configuration corresponds to 
separating $P$ $D$-branes by a distance $L$  from a stack of $N+P$ coincident $D3$-branes and introducing $k$ fundamental strings stretched between the two stacks of branes,  in the limit $L\rightarrow \infty$. 
We can exactly integrate out  the degrees of freedom introduced by the extra $P$ $D$-branes  from the low energy effective field theory describing this configuration and  show that the net effect is to insert into the $U(N)$ ${\cal N}=4$ SYM path integral a Wilson loop operator with the expected representation, thus explicitly confirming the proposal in \GomisSB.

The plan of the rest of this note is as follows. In section $2$ we show that a single $D3$-brane in AdS$_5\times$S$^5$ with $k$ units of fundamental string charge correponds to a half-BPS Wilson loop in the $k$-th symmetric representation of $U(N)$. This is shown by studying in a certain infinite mass limit the Coulomb branch of ${\cal N}=4$ SYM in the presence of $k$ W-bosons. In section $3$ this result is generalized to arbitrary representations and reproduces the proposal in \GomisSB.  Some details of the infinite mass limit of the Coulomb branch of ${\cal N}=4$ SYM are relegated to the  Appendix.

\newsec{A $D3_k$-brane as a Wilson loop in the $k$-th symmetric representation}

In \GomisSB, it was argued that the    $D3_k$-brane solution  in AdS$_5\times$S$^5$ of \ReyIK\DrukkerKX\ corresponds to a half-BPS Wilson loop operator labeled by the following Young tableau:
\medskip
\noindent
\centerline{\young(12\cdot\cdot\cdot\cdot k).
}
\medskip
This solution \ReyIK\DrukkerKX\ has an AdS$_2\times$S$^2$ worldvolume geometry and carries $k$ units of fundamental string charge. The fact that $k$ is arbitrary, that there can be at most $N$ such $D3$-branes in AdS$_5\times$S$^5$, and its proposed relation through bosonization to the defect conformal field theory derived for the $D5_k$-brane\foot{This $D5$-brane, which has an AdS$_2\times$S$^4$ worldvolume geometry and $k\leq N$ units of fundamental string charge, was shown to correspond to a Wilson loop in the $k$-th antisymmetric representation --  a Young tableau with $k$ boxes in one column  --
 by integrating out the degrees of freedom on the $D5$-brane.} led  \GomisSB\ to the abovementioned proposal.  

In this note we show that this proposal is indeed correct by studying a brane configuration in flat space. We   integrate out the physics on the brane and show that the $D$-brane inserts the desired Wilson loop into the ${\cal N}=4$ SYM path integral. This brane configuration can also be studied in the near horizon limit and indeed reproduces the $D3_k$-brane solution of  \ReyIK\DrukkerKX.

A half-BPS Wilson loop of ${\cal N}=4$ SYM in a representation\foot{$R=(R_1,R_2,\ldots,R_N)$, with $R_i\geq R_{i+1}$  labels  a representation of $U(N)$ given by a Young tableau with $R_i$ boxes in the $i$-th row.} $R$
\eqn\half{
W_R=\hbox{Tr}_R P\exp\left(i \int dt\;(A_0+\phi)\right),}
is obtained by adding a static, infinitely massive charged probe to ${\cal N}=4$ SYM. As already shown in \ReyIK\MaldacenaIM (see also
\lref\DrukkerZQ{
  N.~Drukker, D.~J.~Gross and H.~Ooguri,
   ``Wilson loops and minimal surfaces,''
  Phys.\ Rev.\ D {\bf 60}, 125006 (1999)
  [arXiv:hep-th/9904191].
}
\DrukkerZQ), one way of introducing external charges in $U(N)$ ${\cal N}=4$ SYM is to consider a stack of $N+1$ $D3$-branes and going along the Coulomb brach of the gauge theory.

Let's consider the gauge theory on $N+1$ $D3$-branes and break the gauge symmetry down to $U(N)\times U(1)$ by separating one of the branes. In the gauge theory description this corresponds to turning on 
  the following expectation value
\eqn\expect{
\la \phi \ra=\pmatrix{0&0\cr
0&L},}
where $\phi$ is one of the scalar fields of ${\cal N}=4$ SYM, thus breaking the $SO(6)$ R-symmetry of ${\cal N}=4$ SYM down to $SO(5)$.

We are interested in studying the low energy physics of this D-brane configuration in a background where $k$ static fundamental strings are stretched between the two stacks of $D3$-branes:
\ifig\sphera{Two separated stacks of $D3$-branes with $k$ fundamental strings stretched between them.}{\epsfxsize1.5in\epsfbox{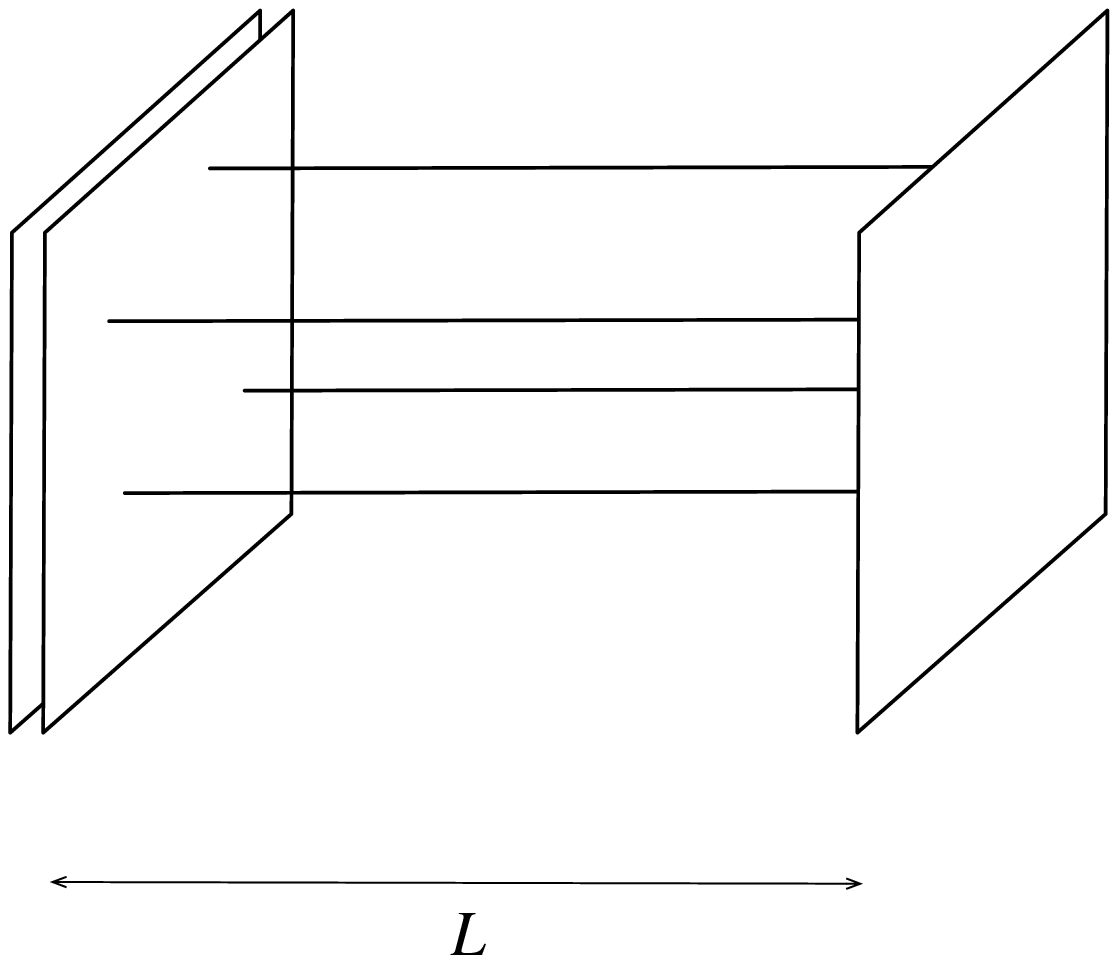}} 

In the gauge theory description, we must study the low energy effective field theory of $U(N+1)$ ${\cal N}=4$  SYM when spontaneously broken to $U(N)\times U(1)$. The presence of $k$ stretched static fundamental strings corresponds to inserting at $t\rightarrow -\infty$ $k$ W-boson creation operators $w^\dagger$ and $k$ W-boson annihilation operators $w$ at $t\rightarrow \infty$. Since we are interested in the limit when the charges are infinitely massive probes, we must study this field theory vacuum  in the limit   $L\rightarrow \infty$. In this limit the $U(1)$ theory completely decouples from the $U(N)$ theory.

Physically, the $L\rightarrow \infty$ limit  can be thought of as a non-relativistic limit. The  dynamics can be conveniently extracted by defining 
\eqn\nonrel{
w={1\over \sqrt{L}}e^{-itL}\chi,}
making the kinetic term for the W-bosons non-relativistic
\lref\GomisBD{
  J.~Gomis and H.~Ooguri,
  ``Non-relativistic closed string theory,''
  J.\ Math.\ Phys.\  {\bf 42}, 3127 (2001)
  [arXiv:hep-th/0009181].
}. 
As shown in the Appendix, the terms in the effective action surviving the limit are given by
\eqn\action{
S=S_{{\cal N}=4}+S_\chi,}
where: 
\eqn\actionw{
S_\chi=\int   i\chi^\dagger \partial_t \chi+\chi^\dagger(A_0+\phi)\chi.}

Therefore, the path integral describing $k$ fundamental strings stretching between the two stacks of $D$-branes in the $L\rightarrow \infty$ limit is given by\foot{The path integral over the $U(N)$ ${\cal N}=4$ SYM is to be performed at the end.}

\eqn\path{
 Z\equiv e^{iS_{{\cal N}=4}}\hskip-5pt\int [D\chi][D\chi^\dagger]\ e^{iS_\chi}
{1\over k!}\hskip-3pt\sum_{i_1,\ldots i_k}\hskip-2pt\chi_{i_1}(\infty)\chi_{i_2}(\infty)\hskip-2pt\ldots\hskip-2pt\chi_{i_k}(\infty)\chi_{i_1}^\dagger(-\infty)\chi_{i_2}^\dagger(-\infty)\hskip-2pt\ldots\hskip-2pt\chi_{i_k}^\dagger(-\infty),}
where $i_l=1,\ldots N$ is a fundamental index of $U(N)$.

From the formula for the W-boson propagator that follows from \actionw\foot{$\theta(t)$ is the Heaviside step function.}
\eqn\prop{
\la \chi_{i}(t_1)\chi_{j}^\dagger(t_2)\ra=\theta(t_1-t_2)\delta_{ij},}
 one can derive the following ``effective" propagator
\eqn\effect{
 \la \chi_{i}(\infty) \chi_j^\dagger(-\infty)\ra_{eff}\equiv \la \exp\left({i  \int dt\ i\chi^\dagger \partial_t \chi+\chi^\dagger(A_0+\phi)\chi}\right)
  \chi_{i}(\infty) \chi_j^\dagger(-\infty)\ra =U_{ij},}
where $U$ is the  holonomy matrix appearing in the half-BPS Wilson loop operator \half:
\eqn\holo{
U=P\exp\left(i \int dt\; (A_0+\phi)\right)\in U(N).}

Using this ``effective" propagator we can now evaluate \path. We must sum over all Wick contractions between the $W$-bosons. Contractions are labeled by a permutation $\omega$ of the symmetric group $S_k$. The path integral \path\ is then given by:
\eqn\fin{
Z=e^{iS_{{\cal N}=4}}\cdot{1\over k!}\sum_{\omega \in S_k} U^{i_1}_{i_{\omega(1)}}\ldots U^{i_k}_{i_{\omega(k)}}.}
Permutations having the same cycle structure upon decomposing a permutation into the product of disjoint cycles give identical contributions in \fin. Since all elements in a given conjugacy class of $S_k$ have the same cycle structure, we can replace the sum over permutations $\omega$ in \fin\ by a sum over conjugacy classes $C(\vec{k})$ of $S_k$. Conjugacy classes of $S_k$ are labeled by partitions of $k$, denoted by $\vec{k}$, so that
\eqn\parti{
k=\sum_{l=1}^k lk_l,}
and each permutation in the conjugacy class has $k_l$ cycles of length $l$. 

Therefore, \fin\ can be written as 
\eqn\conja{
Z=e^{iS_{{\cal N}=4}}\cdot{1\over k!}\sum_{C(\vec{k})}N_{C(\vec{k})}\gamma_{\vec{k}}(U),}
where 
\eqn\winding{
\gamma_{\vec{k}}(U)=\prod_{l=1}^k(\hbox{Tr}U^l)^{k_l},}
and $N_{C(\vec{k})}$ is the number of permutations in the conjugacy class $C(\vec{k})$, which is given by
\eqn\numberof{
N_{C(\vec{k})}={k!\over z_{\vec{k}}},}
with:
\eqn\facto{
z_{\vec{k}}=\prod_{l=1}^k k_l! l^{k_l}.}
Therefore, we are led to
\eqn\correla{
 Z=e^{iS_{{\cal N}=4}}\cdot\sum_{C(\vec{k})}{1\over z_{\vec{k}}}\gamma_{\vec{k}}(U),}
which can also be written (see e.g. 
\lref\groupthe{
W. Fulton and J. Harris, ``Representation Theory", Springer 2000.
  }
\groupthe) as 
 \eqn\finalexpre{
 Z=e^{iS_{{\cal N}=4}}\cdot \hbox{Tr}_{(k,0,\ldots,0)} U,}
 as we wanted to show.
 
 To summarize, we have shown that integrating out the degrees of freedom associated to the single separated $D3$-brane -- when $k$ fundamental strings are stretching between the $D3$-brane and a stack of $N$ $D3$-branes -- inserts a half-BPS Wilson loop operator into the ${\cal N}=4$ SYM path integral in the $k$-th symmetric representation of $U(N)$. 
 
We can now make contact with the $D3_k$-brane  solution \ReyIK\DrukkerKX\ in AdS$_5\times$S$^5$. The solution of the Born-Infeld equations of motion for a single $D3$ brane with $k$ fundamental strings stretched between that brane and a stack of $N$ $D3$-branes was already found in \ReyIK. In this solution,  the $N$ $D3$-branes are replaced by their supergravity background and the other $D3$-brane  with the attached strings as a BION solution
\lref\CallanKZ{
  C.~G.~.~Callan and J.~M.~Maldacena,
  ``Brane dynamics from the Born-Infeld action,''
  Nucl.\ Phys.\ B {\bf 513}, 198 (1998)
  [arXiv:hep-th/9708147].
}
\lref\GibbonsXZ{
  G.~W.~Gibbons,
  ``Born-Infeld particles and Dirichlet p-branes,''
  Nucl.\ Phys.\ B {\bf 514}, 603 (1998)
  [arXiv:hep-th/9709027].
}
\CallanKZ\GibbonsXZ. In the near horizon limit, the $D3$-brane solution in \ReyIK\ indeed becomes the $D3_k$-brane solution in AdS$_5\times$S$^5$. 
 
 Therefore, we have given a microscopic explanation of the 
 identification  
\eqn\pathfin{
D3_k\longleftrightarrow Z=e^{iS_{{\cal N}=4}}\cdot W_{(k,0, \ldots ,0)},}
proposed in \GomisSB. 
 
\newsec{Multiple $D3_k$-branes as Wilson loop in arbitrary representation}

In the previous section, we have shown that a single $D3_k$-brane corresponds to a Wilson loop in the $k$-th symmetric representation. We now show that an arbitrary representation $R$ with $P$ rows in a Young tableau can be realized by considering $P$ $D3$-branes. 

We   consider a stack of $N+P$ $D3$-branes and break the gauge symmetry down to $U(N)\times U(P)$ by separating $P$ of the branes a distance $L$. In the gauge theory description this corresponds to turning on a scalar expectation value as in \expect. We also consider a background of $k$ fundamental strings stretched between the two stacks of branes. 

Therefore, we must study the low energy effective field theory of $U(N+P)$ ${\cal N}=4$ SYM when spontaneously broken to $U(N)\times U(P)$ and in the limit $L\rightarrow \infty$, where the charges become infinitely massive probes\foot{Just as before, the $U(P)$ gauge dynamics completely decouples from the $U(N)$ gauge theory in the $L\rightarrow \infty$ limit.}. The presence of $k$ fundamental strings is realized in the gauge theory by inserting the creation operator of a $k$ W-boson state at $t\rightarrow -\infty$ and the annihilation operator of a $k$ W-boson state  at $t\rightarrow  \infty$. The $k$ W-boson annihilation operator is given by
\eqn\defchi{\Psi(t)=\chi_{i_1}^{I_1}(t)\chi_{i_2}^{I_2}(t)\ldots \chi_{i_k}^{I_k}(t)} and the $k$ W-boson creation operator by $\Psi^\dagger(t)$, where\foot{The W-bosons transform in the $(N,\bar{P})$ representation of the $U(N)\times U(P)$  gauge group, see the Appendix for details.} $i_l=1,\ldots,N$ and $I_l=1,\ldots,P$.

Such a $k$ W-boson state transforms under $U(N)$ and $U(P)$ as a sum over representations with $k$ boxes in a Young tableau. In order to project to 
a specific representation $R$ we can apply to the  $k$ W-boson annihilation operator \defchi\ the following projection operator 
\eqn\proel{P_\alpha^R={d_R\over k!}\sum_{\sigma\in S_k}D_{\alpha\alpha}^R(\sigma)\sigma}  
where $R=(n_1,n_2,\ldots ,n_P)$, with $k=\sum_i n_i$,  labels an irreducible  representation of both\foot{There is a natural action of $S_k$, $U(N)$ and $U(P)$ on $\Psi(t)$.  The projected operator in fact transforms in the same representation $R$ for  both $S_k$, $U(N)$ and $U(P)$  groups (see e.g.
\lref\group{J-Q Chen, J. ping and F. Wang, ``Group Representation for Physicists," World Scientific.}
\group). The representations of the unitary and symmetric groups are both labeled by the same Young tableau $R=(n_1,n_2,\ldots ,n_P)$.} $S_k$, $U(N)$ and $U(P)$. $D_{\alpha\beta}^R(\sigma)$ is the representation matrix for the permutation $\sigma$ in the  representation $R$, $d_R$ is the dimension of the representation $R$ of $S_k$ and $\alpha,\beta=1,\ldots,d_R$. Therefore, the operator  \eqn\wstate{\Psi_{\alpha}^{R}(t)=P_\alpha^R\Psi={d_R\over k!}\sum_{\sigma\in S_k}D_{\alpha\alpha}^R(\sigma)\chi_{i_1}^{I_{\sigma(1)}}(t)\chi_{i_2}^{I_{\sigma(2)}}(t)\ldots \chi_{i_k}^{I_{\sigma(k)}}(t)} 
describes a $k$ W-boson state
transforming in the irreducible representation $R$ of $S_k$, $U(N)$ and $U(P)$.

The path integral to perform, representing our brane configuration with  $k$ fundamental strings stretching between the two stacks of $D$-branes, in the $L\rightarrow \infty$ limit is given by\foot{To avoid cluttering the formulas, the sum over $U(N)$ and $U(P)$ indices is not explicitly written  throughout the rest of this note.}
\eqn\pathnew{
 Z= e^{iS_{{\cal N}=4}}\hskip-5pt\int [D\chi][D\chi^\dagger]\ e^{iS_\chi}
 \sum_{\alpha=1}^{d_R}\Psi_\alpha^{ R}(\infty)\Psi_\alpha^{\dagger R}(-\infty),}
where $S_{\chi}$ is the straightforward generalization of \actionw\ when the gauge group is $U(N)\times U(P)$. 

The ``effective" propagator for the W-bosons is now 
\eqn\effectnon{
 \la \chi_{i}^I(\infty) \chi_j^{J\dagger}(-\infty)\ra_{eff}\equiv \la \exp\left({i  \int dt\ i\chi^\dagger \partial_t \chi+\chi^\dagger(A_0+\phi)\chi}\right)
  \chi^I_{i}(\infty) \chi_j^{J\dagger}(-\infty)\ra =U_{ij}\delta^{IJ},}
with $U$  given in \holo. The sum over all Wick contractions in \pathnew\ gives:
\eqn\contractnon{
Z=e^{iS_{{\cal N}=4}}\left({d_R\over k!}\right)^2\sum_{\alpha=1}^{d_R}
\sum_{\sigma,\tau,\omega \in S_k}D_{\alpha\alpha}^R(\sigma)D_{\alpha\alpha}^R(\tau)U^{i_1}_{i_{\omega(1)}}\ldots U^{i_k}_{i_{\omega(k)}}\delta^{I_{\sigma(1)}}_{I_{\omega\tau(1)}}\ldots \delta^{I_{\sigma(k)}}_{I_{\omega\tau(k)}}.}
By appropriate change of variables, this can be simplified to
\eqn\simply{
Z= e^{iS_{{\cal N}=4}}\left({d_R\over k!}\right)^2\sum_{\alpha=1}^{d_R}
\sum_{\sigma,\tau,\omega \in S_k}D_{\alpha\alpha}^R(\sigma)D_{\alpha\alpha}^R(\tau)U^{i_1}_{i_{\omega(1)}}\ldots U^{i_k}_{i_{\omega(k)}}P^{C(\sigma^{-1}\omega\tau)},}
where $C(\sigma)$ is the number of disjoint cycles in the permutation $\sigma$ and:
\eqn\deltas{
P^{C(\sigma^{-1}\omega\tau)}=\sum_{I_1,\ldots,I_k}
\delta^{I_1}_{I_{\sigma^{-1}\omega\tau(1)}}\ldots \delta^{I_k}_{I_{\sigma^{-1}\omega\tau(k)}}.}

We proceed\foot{The paper 
\lref\CorleyZK{
  S.~Corley, A.~Jevicki and S.~Ramgoolam,
  ``Exact correlators of giant gravitons from dual N = 4 SYM theory,''
  Adv.\ Theor.\ Math.\ Phys.\  {\bf 5}, 809 (2002)
  [arXiv:hep-th/0111222].
}
\CorleyZK\ has a useful compilation of useful formulas relevant for this paper.} by introducing $\delta(\rho)$, an element in the group algebra, 
which takes the value $1$ when the argument is the identity permutation and $0$ when the argument is any other permutation.  This allows \simply\ to be written as: 
\eqn\propdel{e^{iS_{{\cal N}=4}}\left({d_R\over k!}\right)^2\sum_{\alpha=1}^{d_R}\sum_{\sigma,\tau,\omega,\rho\in S_k}D_{\alpha\alpha}^R(\sigma)D_{\alpha\alpha}^R(\tau)U^{i_1}_{i_{\omega(1)}}\ldots U^{i_k}_{i_{\omega(k)}}P^{C(\rho)}\delta({\rho^{-1}\sigma^{-1}\omega\tau}).}
Summing over $\tau$ yields
\eqn\propsum{e^{iS_{{\cal N}=4}}\left({d_R\over k!}\right)^2\sum_{\alpha=1}^{d_R}\sum_{\sigma,\omega\in S_k}D_{\alpha\alpha}^R(\sigma)D_{\alpha\alpha}^R(\omega^{-1}\sigma\sum_{\rho\in S_k}\rho P^{C(\rho)})U^{i_1}_{i_{\omega(1)}}\ldots U^{i_k}_{i_{\omega(k)}}.}
Since ${\cal C}=\sum_{\rho\in S_k}\rho P^{C(\rho)}$ commutes with all elements in the group algebra, we can use the identity
$D_{\alpha\alpha}^R({\cal C}\sigma)={1\over d_R}D_{\alpha\alpha}^R(\sigma)\chi_R({\cal C})$, where $\chi_R({\cal C})=\sum_{\alpha=1}^{d_R}D_{\alpha\alpha}^R({\cal C})$ is the character of $S_k$ in the representation $R$ for ${\cal C}$. Therefore, \propsum\ reduces to 
\eqn\propdim{e^{iS_{{\cal N}=4}}{d_R\over k!} Dim_P(R)\sum_{\alpha=1}^{d_R}\sum_{\sigma,\omega\in S_k}D_{\alpha\alpha}^R(\sigma)D_{\alpha\alpha}^R(\omega^{-1}\sigma)U^{i_1}_{i_{\omega(1)}}\ldots U^{i_k}_{i_{\omega(k)}},}
where \eqn\dim{Dim_P(R)={1\over k!}\sum_{\sigma\in S_k}\chi_R(\sigma)P^{C(\sigma)}} is the dimension of the irreducible representation $R$ of   $U(P)$. By using the relation satisfied by the fusion of representation matrices
\eqn\matc{\sum_{\sigma\in S_k}D_{\alpha\alpha}^R(\sigma)D_{\alpha\alpha}^R(\omega^{-1}\sigma)={k!\over d_R}D_{\alpha\alpha}^R(\omega^{-1})}
we are arrive at:
\eqn\propdim{
Z=e^{iS_{{\cal N}=4}}Dim_P(R)\sum_{\omega \in S_k}\chi_R(\omega)U^{i_1}_{i_{\omega(1)}}\ldots U^{i_k}_{i_{\omega(k)}}.}

Finally, we use the Frobenius character formula (see e.g. \groupthe), which relates the trace of a matrix $U$ in an arbitrary representation $R=(n_1,n_2,\ldots,n_P)$ of $U(N)$ to the trace in the fundamental representation
\eqn\car{\hbox{Tr}_R(U)={1\over k!}\sum_{\omega \in S_k}\sum_{i_1,\ldots i_k}\chi_R(\omega)U^{i_1}_{i_{\omega(1)}}\ldots U^{i_k}_{i_{\omega(k)}}}
to show that the final result of the path integral is  
\eqn\fino{
Z=e^{iS_{{\cal N}=4}}\cdot k!\; Dim_R(M)\hbox{Tr}_R(U),}
the insertion of a half-BPS Wilson loop in the representation $R$.

In the near horizon limit, when the $N$ D3-branes are replaced by their near horizon geometry, the $P$ D3-branes with the array of stretched fundamental strings  labeled by $R=(n_1,n_2,\ldots,n_P)$ become the brane configuration $(D3_{n_1},D3_{n_2},\ldots,D3_{n_P})$  in AdS$_5\times$S$^5$, thus arriving at  the identification\foot{We  can  trivially reabsorb the overall constant in \fino\ in the normalization of $\Psi$.}
\eqn\pathfinfinfin{
(D3_{n_1},\ldots, D3_{n_P})\longleftrightarrow Z=e^{iS_{{\cal N}=4}}\cdot W_{(n_1,\ldots,n_P,0,\dots,0)}}
in \GomisSB.

\bigbreak\bigskip\bigskip\centerline{{\bf Acknowledgements}}\nobreak
We would like to thank F. Cachazo, L. Freidel and  C. R\"omelsberger for enjoyable discussions. Research at the Perimeter Institute is supported in part by funds from NSERC of Canada and by MEDT of Ontario. We also acknowledge further  support by an NSERC Discovery Grant.

\appendix{A}{Gauge Theory Along Coulomb Branch}

The low energy dynamics of  a stack of $N+P$  coincident $D$3-branes is described by four dimensional ${\cal N}=4$  SYM with $U(N+P)$ gauge group.  The spectrum of the theory includes  a vector field $\hat{A}_{\mu}$, six scalar fields  $\hat{\Phi}_i$ and a  ten dimensional Majorana-Weyl spinor $\hat{\Psi}$.    The action is given by 
\eqn\sym{\hat{S}_{{\cal N}=4}={1\over 2  g_{YM}^2}\int    {\rm Tr} \left (-{1\over 2}\hat{F}_{\mu\nu}^2 -(\hat{D}_{\mu}\hat{\Phi}_i)^2+{1\over 2} [\hat{\Phi}_i ,\hat{\Phi}_j]^2 -i\hat{\bar{\Psi}} \Gamma^{\mu} \hat{D}_{\mu}\hat{\Psi} - \hat{\bar{\Psi}} \Gamma^{i} [ \hat{\Phi}_i ,\hat{\Psi}] \right ),}
where each field is in the adjoint of the gauge group $U(N+P)$.  We use  real ten dimensional gamma matrices  $\Gamma^i$ and $\Gamma^\mu$ and we choose $\Gamma^0$ as charge conjugation  matrix.   Thus, the Majorana-Weyl spinor  $\lambda$ has 16 real components and  $\bar{\lambda}=\lambda^T\Gamma^0$.

Now we separate a stack of  $P$ branes from the remaining stack of $N$ branes,  i.e. we   give  a non trivial vacuum expectation value to the scalar fields. Without lost of generality,   we take
\eqn\vac{<\hat{\Phi}_9>=\pmatrix{0&0\cr 0&LI_P},} 
 where $I_P$ is the  $P\times P$ unit matrix and $L$ is a constant  with the dimensions of   mass.     To expand the action around this vacuum, we first  define the fields as 
\eqn\red{\hat{A}_\mu=\pmatrix{A_\mu& \omega_\mu\cr \omega_\mu^\dagger & \tilde{A}_\mu}\qquad \hat{\Phi}_i=\pmatrix{\Phi_i &  \omega_i\cr \omega_i^\dagger & \delta_{i9}LI_P+\tilde{\Phi}_i }\qquad \hat{\Psi}=\pmatrix{\Psi & \theta \cr \theta^\dagger & \tilde{\Psi}},} 
 where $A_\mu$, $\Phi_i$ and  $\Psi$ transform in the adjoint representation of $U(N)$ and  $\tilde{A}_\mu$,  $\tilde{\Phi}_i$ and  $\tilde{\Psi}$ transform in the adjoint representation of $U(P)$.     $\omega_\mu$, $\omega_i$ and $\theta$  are W-bosons fields and    transform in the $(N,\bar{P})$ representation  of the gauge group    $U(N)\times U(P)$.     
 
 The action becomes:
 \eqn\sepa{\hat{S}_{{\cal N}=4}^{L}=S_{{\cal N}=4}+\tilde{S}_{{\cal N}=4} + S_{W}+S_{interactions}.}
$S_{{\cal N}=4}$ and  $\tilde{S}_{{\cal N}=4}$ are the actions  for the effective field theories living on the two stacks of branes,  i.e.  four dimensional  ${\cal N}=4$ SYM with gauge group respectively  $U(N)$ and $U(P)$.  
$S_{W}$  is the   action for the W-bosons and their superpartners        
\eqn\wquad{ S_{W}=\int    {\rm Tr} \bigg(-{1\over 2}f_{\mu\nu}^\dagger f^{\mu\nu} -  L^2 \omega_\mu^\dagger\omega^{\mu} 
 - \sum_{i=4}^{9}\partial^\mu \omega_i^\dagger\partial_\mu\omega_i -    L^2  \sum_{k=4}^{8} \omega_k^\dagger \omega_k   - i\bar{\theta}^\dagger\Gamma^\mu \partial_\mu \theta+ L\bar{\theta}^\dagger\Gamma^9\theta+\ldots \bigg),}    
 where  $f_{\mu\nu}=\partial_\mu \omega_\nu-\partial_\nu\omega_\mu$  
 and $\ldots$ denote terms fourth order in the W-boson fields.
$S_{interactions}$  is the action  describing the interactions between the W-bosons and the fields living on the two stacks of branes.   It includes terms of the third and fourth order in the fields.

We are interested  in the limit where the two stacks  of branes are infinitely separated, i.e. in the limit where $L\rightarrow \infty$.   From the quadratic action \wquad\  we see that the W-bosons $\omega_m$ with $m=1,\ldots 8$ and the fermions  become infinitely massive. Taking  the infinite mass limit of a relativistic massive field,   corresponds to considering the non-relativistic limit.    The 
surviving dynamics in the limit can be explicitly extracted by making the following redefinition:
\eqn\resca{\omega_m={1\over \sqrt{L}}e^{-itL}\chi_m\qquad{\rm where }\qquad m=1,\ldots,8.} 
For the W-bosons superpartners, which also become infinitely massive,  we first define
\eqn\xipm{\Gamma_{09}\theta_{\pm}=\pm \theta_{\pm},}  where this projection must be understood in the spinors space.  To extract the physics in the limit we make the following rescaling:
\eqn\fresca{\theta=\theta_++\theta_-=e^{-iLt}\xi_++e^{-iLt}\xi_-.}
 Considering \resca\  and \fresca\  and then taking the infinite mass limit $L\rightarrow \infty$ the W-boson  action \wquad\   reduces to 
 \eqn\wquadnr{S_{W}^{NR}=\int   {\rm Tr} \left(i\sum_{m=1}^{8}\chi_m^\dagger \partial_t\chi_m   +i (\xi_+^T)^\dagger \partial_t \xi_+  \right)} 
 where the transposition is in the space of fermions and the hermitian conjugation  is in the matrix space. The $\xi_-$ fermions become infinitely massive and decouple from the theory, as expected, since there are no antiparticles in the non-relativistic limit.
 
The interaction  action $S_{interaction}$ in \sepa\ is now given by    
\eqn\intnr{\eqalign{S_{interactions}^{NR}=\int   \sum_{m=1}^{8} {\rm Tr} \bigg( \chi_m^\dagger( A_0 +\Phi_9)\chi_m -  \chi_m^\dagger   \chi_m(\tilde{A}_0+\tilde{\Phi}_9) + \cr +  \xi_m^\dagger( A_0 +\Phi_9)\xi_m -  \xi_m^\dagger\xi_m(\tilde{A}_0 +\tilde{\Phi}_9)\bigg)}}
 where  $\xi_m$ $(m=1,\ldots 8) $ are the spinor components of $\xi_+$.  All higher order terms in \sepa\ vanish in the $L\rightarrow \infty$ limit. Note that, in this limit the dynamics of the $U(P)$ gauge theory effectively decouples from the $U(N)$ gauge theory.

 Therefore, we can then 
 write the action describing the coupling of the W-bosons to $U(N)$ ${\cal N}=4$ SYM as\foot{There is decoupled contribution for the $U(P)$ gauge theory which does not talk to  $U(N)$ SYM.} 
 \eqn\twolag{S=S_{{\cal N}=4}+\sum_{m=1}^{8} S_{m},  }
 where  $S_{{\cal N}=4}$ is the action of ${\cal N}=4$ SYM with gauge group $U(N)$ while  $S_{m}$ with $m=1,\ldots 8$ is the action for one of the eight non-relativistic supersymmetric W-bosons 
  \eqn\com{\eqalign{S_{m}=\int  [ (\chi_m^\dagger)^{I}_i \Delta_{ij}^{IJ}(\chi_m)_j^J+ (\xi_m^\dagger)^{I}_i \Delta_{ij}^{IJ}(\xi_m)_j^J]},} 
  where 
  \eqn\deltaop{\Delta_{ij}^{IJ}=i\delta_{ij}\delta^{IJ}\partial_t+(A_{0}+\Phi_9)_{ij}\delta^{IJ},}
which is what we have used in the main text.  
  
  We note that integrating out the degrees of freedom associated to the W-bosons, without any insertions,  we get 
 \eqn\zvacuum{\eqalign{Z=&\int\prod_{i=1}^{8}([D\chi_m][D\chi_m^\dagger][D\xi_m][D\xi_m^\dagger])e^{iS} \cr=&e^{iS_{{\cal N}=4}}{({\rm det}  \Delta)^{n_F}\over ({\rm det}  \Delta)^{n_B}}  \cr  =&e^{iS_{{\cal N}=4}},}}
 where in the last step we used that $n_F=n_B$. Note that we  recover the
 expected result that the metric in the Coulomb branch of ${\cal N}=4$ gets no corrections upon integrating out the massive modes.

\listrefs
\bye

%% file: youngtab.tex
\catcode`\@11\relax
\newif\ify@autoscale \y@autoscaletrue \def\Yautoscale#1{\ifnum #1=0
  \y@autoscalefalse\else\y@autoscaletrue\fi}
\newdimen\y@b@xdim
\newdimen\y@boxdim \y@boxdim=13pt
\def\Yboxdim#1{\y@autoscalefalse\y@boxdim=#1}
\newdimen\y@linethick    \y@linethick=.3pt
\def\Ylinethick#1{\y@linethick=#1}
\newskip\y@interspace \y@interspace=0ex plus 0.3ex
\def\Yinterspace#1{\y@interspace=#1}
\newif\ify@vcenter   \y@vcenterfalse
\def\Yvcentermath#1{\ifnum #1=0 \y@vcenterfalse\else\y@vcentertrue\fi}
\newif\ify@stdtext   \y@stdtextfalse
\def\Ystdtext#1{\ifnum #1=0 \y@stdtextfalse\else\y@stdtexttrue\fi}
\newif\ify@enable@skew   \y@enable@skewfalse
\expandafter\ifx\csname enableskew\endcsname\relax
 \y@enable@skewfalse \else \y@enable@skewtrue\fi
\def\y@vr{\vrule height0.8\y@b@xdim width\y@linethick depth 0.2\y@b@xdim}
\def\y@emptybox{\y@vr\hbox to \y@b@xdim{\hfil}}
\ify@enable@skew
 \def\y@abcbox#1{\if :#1\else
   \y@vr\hbox to \y@b@xdim{\hfil#1\hfil}\fi}
 \def\y@mathabcbox#1{\if :#1\else
   \y@vr\hbox to \y@b@xdim{\hfil$#1$\hfil}\fi}
\else
 \def\y@abcbox#1{\y@vr\hbox to \y@b@xdim{\hfil#1\hfil}}
 \def\y@mathabcbox#1{\y@vr\hbox to \y@b@xdim{\hfil$#1$\hfil}}
\fi
\def\y@setdim{%
  \ify@autoscale%
   \ifvoid1\else\typeout{Package youngtab: box1 not free! Expect an
     error!}\fi%
   \setbox1=\hbox{A}\y@b@xdim=1.6\ht1 \setbox1=\hbox{}\box1%
  \else\y@b@xdim=\y@boxdim \advance\y@b@xdim by -2\y@linethick
  \fi}
\newcount\y@counter
\newif\ify@islastarg
\def\y@lastargtest#1,#2 {\if\space #2 \y@islastargtrue
  \else\y@islastargfalse\fi}
\def\y@emptyboxes#1{\y@counter=#1\loop\ifnum\y@counter>0
  \advance\y@counter by -1 \y@emptybox\repeat}
\def\y@nelineemptyboxes#1{%
  \vbox{%
    \hrule height\y@linethick%
    \hbox{\y@emptyboxes{#1}\y@vr}
    \hrule height\y@linethick}\vskip-\y@linethick}
\def\yng(#1){%
  \y@setdim%
  \hskip\y@interspace%
  \ifmmode\ify@vcenter\vcenter\fi\fi{%
  \y@lastargtest#1,
  \vbox{\offinterlineskip
    \ify@islastarg
     \y@nelineemptyboxes{#1}
    \else
     \y@ungempty(#1)
    \fi}}\hskip\y@interspace}
\def\y@ungempty(#1,#2){%
  \y@nelineemptyboxes{#1}
  \y@lastargtest#2,
  \ify@islastarg
   \y@nelineemptyboxes{#2}
  \else
   \y@ungempty(#2)
  \fi}
\def\y@nelettertest#1#2. {\if\space #2 \y@islastargtrue
  \else\y@islastargfalse\fi}
\def\y@abcboxes#1#2.{%
  \ify@stdtext\y@abcbox#1\else\y@mathabcbox#1\fi%
  \y@nelettertest #2.
  \ify@islastarg\unskip%
   \ify@stdtext\y@abcbox{#2}\else\y@mathabcbox{#2}\fi%
  \else\y@abcboxes#2.\fi}
 \newdimen\y@full@b@xdim
 \newcount\y@m@veright@cnt
\ify@enable@skew
 \def\y@get@m@veright@cnt#1#2.{%
   \if :#1 \advance\y@m@veright@cnt by 1\y@get@m@veright@cnt#2.\fi}
 \let\y@setdim@=\y@setdim
 \def\y@setdim{%
   \y@setdim@ \y@full@b@xdim=\y@b@xdim
   \advance\y@full@b@xdim by 1\y@linethick}
 \def\y@m@veright@ifskew#1{
   \y@m@veright@cnt=0 \y@get@m@veright@cnt#1.
   \moveright \y@m@veright@cnt\y@full@b@xdim}
\else
 \def\y@m@veright@ifskew#1{}
\fi
\def\y@nelineabcboxes#1{%
  \y@nelettertest #1.
  \ify@islastarg
   \y@m@veright@ifskew{#1}
    \vbox{
      \hrule height\y@linethick%
      \hbox{\ify@stdtext\y@abcbox#1\else\y@mathabcbox#1\fi\y@vr}
      \hrule height\y@linethick}\vskip-\y@linethick
  \else
   \y@m@veright@ifskew{#1}
    \vbox{
      \hrule height\y@linethick%
      \hbox{\y@abcboxes #1.\y@vr}%
      \hrule height\y@linethick}\vskip-\y@linethick
  \fi}
\def\young(#1){%
  \y@setdim%
  \hskip\y@interspace%
  \y@lastargtest#1,
  \ifmmode\ify@vcenter\vcenter\fi\fi{%
  \vbox{\offinterlineskip
    \ify@islastarg\y@nelineabcboxes{#1}%
    \else\y@ungabc(#1)%
    \fi}}\hskip\y@interspace}
\def\y@ungabc(#1,#2){%
  \y@nelineabcboxes{#1}%
  \y@lastargtest#2,
  \ify@islastarg\y@nelineabcboxes{#2}%
  \else\y@ungabc(#2)%
  \fi}
\catcode`\@12\relax
 